\documentclass[12pt,preprint]{aastex}

\sfcode`A=1000 \sfcode`B=1000 \sfcode`C=1000 \sfcode`D=1000
\sfcode`E=1000 \sfcode`F=1000 \sfcode`G=1000 \sfcode`H=1000
\sfcode`I=1000 \sfcode`J=1000 \sfcode`K=1000 \sfcode`L=1000
\sfcode`M=1000 \sfcode`N=1000 \sfcode`O=1000 \sfcode`P=1000
\sfcode`Q=1000 \sfcode`R=1000 \sfcode`S=1000 \sfcode`T=1000
\sfcode`U=1000 \sfcode`V=1000 \sfcode`W=1000 \sfcode`X=1000
\sfcode`Y=1000 \sfcode`Z=1000

\newcommand{\Lsol}{L$_{\odot}$}
\newcommand{\0}{\phantom{0}}  
\newcommand{\no}{\nodata} 

\slugcomment{Accepted by AJ}
\shorttitle{Radio Sources in EGS}
\shortauthors{Willner et al.}

\begin{document}


\title{Mid-infrared Identification of 6~cm Radio Source Counterparts
in the Extended Groth Strip}


\author{
S.~P.~Willner,\altaffilmark{1}
A.~L.~Coil,\altaffilmark{2,3}
W.~M.~Goss,\altaffilmark{4}
M.~L.~N.~Ashby,\altaffilmark{1}
P.~Barmby,\altaffilmark{1}
J.-S.~Huang,\altaffilmark{1}
R.~Ivison,\altaffilmark{5,6}
D.~C.~Koo,\altaffilmark{7}
E.~Egami\altaffilmark{8}
\&
Satoshi Miyazaki\altaffilmark{9}
}
\altaffiltext{1}{Harvard-Smithsonian Center for Astrophysics, 60 Garden Street,
Cambridge, MA 02138}
\altaffiltext{2}{Department of Astronomy, University of California,
Berkeley, CA 94720--3411}
\altaffiltext{3}{Hubble Fellow, Steward Observatory, University of
  Arizona, Tucson, AZ 85721}
\altaffiltext{4}{National Radio Astronomy Observatory, P.O.~Box 0, 
1003 Lopezville Road, Socorro, NM 87801}
\altaffiltext{5}{Astronomy Technology Centre, Royal Observatory,
Blackford Hill, Edinburgh EH9 3HJ UK}
\altaffiltext{6}{Institute for Astronomy, University of Edinburgh,
Blackford Hill, Edinburgh EH9 3HJ UK}
\altaffiltext{7}{UCO/Lick Observatory, Dept. of Astronomy \&
  Astrophysics, Univ. of California, Santa Cruz, CA 95064}
\altaffiltext{8}{Steward Observatory, University of Arizona, 933
  North Cherry Avenue, Tuscon, AZ 85721}
\altaffiltext{9}{Subaru Telescope, National Astronomical Observatory
  of Japan, 650 North A'ohoku Place, Hilo, HI 96720}


\begin{abstract}
A new 6~cm survey of almost 0.6~deg$^2$ to a limit of 0.55~mJy/beam
(10$\sigma$) finds 37 isolated radio sources and 7 radio source pairs
(not necessarily physical companions).  IRAC counterparts are
identified for at least 92\% of the radio sources
within the area of deep IRAC coverage, which includes
31 isolated sources and 6 pairs.  This contrasts with an
identification rate of $<$74\% to $R<23.95$ in visible light.  Eight
of the IRAC galaxies have power law spectral energy distributions,
implying that the mid-infrared emission comes from a powerful
AGN. The remaining 26 IRAC galaxies show stellar emission in the
mid-infrared, probably in most of these galaxies because the stellar
emission is bright enough to outshine an underlying AGN.  The
infrared colors suggest that the majority of these galaxies are
bulge-dominated and have redshifts $0.5\la z \la 1$.  Visible spectra
from the DEEP2 redshift survey,
available for 11 galaxies, are consistent with this suggestion.  The
IRAC galaxies fall into two distinct groups in a color-magnitude
diagram, one group (the ``stripe'') includes all the AGN.  The
other group (the ``blue clump'') has
blue 3.6 to 8~\micron\ colors and a small range of 8~\micron\
magnitudes.  This separation should be useful in classifying galaxies
found in other radio surveys.

\end{abstract}





\section{Introduction}

Radio observations are an excellent way to identify star-forming 
galaxies and active galactic nuclei (AGN), being subject to fewer 
selection effects than optical surveys (e.g.\ obscuration; spectral 
line contamination). Until recently a large fraction of the most 
distant known galaxies had first been detected in radio surveys, their 
nature betrayed by steep radio spectra and faint optical/infrared 
(opt/IR) counterparts \citep[e.g.,][]{vanB1999}. Indeed, the 
discovery by \citet{Minkowski1960} that 3C~295 --- an object we shall meet 
later in this paper --- lies at $z=0.461$ (at that time the most 
distant known object) was due to its initial detection at 159~MHz and 
a photometric redshift prediction by \citet{Baum1957} of the kind that is 
now commonplace \citep[e.g.,][]{Blake2005}.

Wide, low-frequency surveys such as 3CR tend to yield associations
with bright ellipticals (radio galaxies) and a smattering of radio
quasars and BL Lacs \citep[e.g.,][]{Spinrad1985}, whereas at higher
frequencies (e.g.\ 5~GHz) radio quasars become more common. At lower
flux densities, around 1~mJy, the source counts steepen: evidence of
either primeval radio galaxies \citep{Windhorst1986} or that the
star-forming galaxies play an increasingly important role
\citep{vanderLaan1983}. Indeed, for low radio luminosities
($L_{1.4GHz}\le 10^{23}$~W~Hz$^{-1}$) \citet{Yun1999} argue that the
fraction of AGN in radio-selected samples drops to around 10\%, as
revealed via an excess of radio emission above that predicted by the
far-IR/radio correlation \citep[e.g.,][]{Helou1985}.  All of these
AGN also reveal their nuclear activity in the mid-infrared or X-ray
wavebands.

Determining the nature of faint radio sources is non-trivial. Even
the first step --- obtaining optical or IR identifications --- can be
time consuming, a task that ultimately yields little but a slit
position for spectroscopic follow up and a very basic measure of
stellar mass and star-formation rate, both subject to obscuration by
dust for galaxies at high redshift.

This paper presents moderately-deep radio observations of the
Extended Groth Strip (EGS) obtained at a wavelength of 6~cm using the
Very Large Array. Deeper observations at 20~cm have been obtained
\citep{Ivison2006}, but this paper gives an overview of
the counterparts of the bright population, those sources we expect to
be dominated by AGN.  For distant galaxies, the rest-frame near-IR
observations provided by {\em Spitzer} enable relatively easy
identifications and the best possible measure of stellar mass
\citep[e.g.,][]{Bell2001} and hence of the likely evolutionary state
of a radio galaxy.

Throughout this paper, magnitudes are in the Vega system, and the 
notation [$w$] means the Vega magnitude at wavelength $w$.

\section{VLA Observations}

The EGS is a region of sky that includes the original HST WFPC2 Groth
Strip Survey.  (See \citealt{Vogt2005} for details of the original
HST survey.)  The EGS is centered near J2000 RA=14$^h$ 19$^m$ 00$^s$
Dec=52\degr\ 50\arcmin\ 00\arcsec\ and covers roughly 2\degr\ by
0\fdg25 at a position angle of 50\degr.  Observations within the EGS
(many now being labelled ``AEGIS'') include deep optical imaging,
deep {\it Spitzer} mid-IR imaging \citep{Fazio2004, Barmby2006}, and
X-ray observations with both Chandra \citep{egs_cxo} and XMM Newton
\citep{was03}.  There is also Keck spectroscopy from the
DEEP2 Redshift Survey \citep{Davis2003} covering the redshift range
$0.1<z<1.4$ for normal galaxy types.  Not all observations cover the
entire strip area.  In particular, the {\it Spitzer} observations
include only a 10\arcmin-wide strip centered within the EGS region.
\citet{Davis2006} summarize many of the data sets.

We chose to image the EGS with the VLA at 6~cm instead of 20~cm
because the bright radio source 3C~295 is located nearby at RA=14$^h$
09$^m$ 33\fs49 and Dec=52\degr\ 26\arcmin\ 13\farcs0 (J2000), only
8\farcm5 from the 
southwest corner of the Spitzer coverage in the EGS.  At 6~cm,
contamination levels are much lower than at 20~cm because 3C~295 is a
factor of 3.5 times fainter and 3.3 times farther away in terms of
the primary beam size.  The shorter observing wavelength makes this
survey relatively more sensitive to flat-spectrum radio sources than
a 20~cm survey would be.

The observations were obtained at the Very Large Array (VLA) in BnA
configuration for a total of 19 hours over 3 days from 2003 October
11--13 (program AW615).  This configuration gives wide coverage on
the sky with angular resolution similar to that of the IRAC imaging
data, as required for identifying IRAC counterparts to the radio
sources.    At 4.8~GHz the VLA
antennas have a primary beam with a FHWM of $9'$.  The mapping grid
contained 74 pointings, spaced $5'$ apart, providing roughly uniform
sensitivity beyond the full Spitzer area in the EGS.  Each pointing
center was observed for 15 minutes.  The observations were carried
out in continuum mode with two intermediate frequency (IF) bands,
each 50 MHz wide, centered at 4885 and 4835~MHz. The point source
VCS2 J1400+6210 (=4C~62.22) was observed for 2 minutes every 17
minutes to provide phase calibrations.  Phase stability was
sufficient to give astrometric accuracy better than 0\farcs1 rms.  We
observed the flux calibrators 3C~286 for 12 minutes on October 12 and
3C~48 for 15 minutes on October~13.  No flux calibrator was observed
on October~11.

The data were reduced using the AIPS software package.  We used TVFLG
to ignore bad visibility data with discrepant amplitude values.  Flux
calibrations were derived from observations of 3C~286 and 3C~48, which
gave 1.71~Jy for the flux density of the phase calibrator.
Measurements with the two primary calibrators agreed within 2\%,
which we take as an estimate of the flux calibration uncertainty.  We
then used 4C~62.22 to provide phase calibrations and
amplitudes for the October~11 data.  Images containing bright
($>$10~mJy/beam) point sources were self-calibrated.


For each of the 74 pointings we used the IMAGR task in AIPS to create
a 2048x2048 pixel image with a pixel scale of 0\farcs4 per pixel.
ROBUST was set to 0, which is an intermediate case between uniform
and natural weighting of the sparsely-sampled UV points, and UVTAPER
was set to 170 k$\lambda$, effectively de-weighting the A-array
elements and reducing the elongation of the synthesized beam.  To
avoid clean bias we CLEANed the images to a flux level of
260~$\mu$Jy/beam, corresponding to $\sim$4$\sigma$~rms, which typically
took 100--200 CLEAN iterations.  The 1$\sigma$~rms in these images is
$\sim$60--70~$\mu$Jy/beam.  Each image is 13\farcm65 on a side, larger
than the FWHM of the primary beam. The synthesized beam is
approximately 1\farcs0 by 1\farcs5 at a PA of 25\degr, but it varies
slightly from pointing to pointing. This size is similar to the
IRAC FWHM of 1\farcs7--2\farcs0 \citep{irac}. We also imaged a small region
at the location of 3C~295 and found no residual flux
with an rms noise of 67~$\mu$Jy/beam.  Thus any artifacts
from 3C~295 should be below the 1$\sigma$ level.

For each quarter of the length of the strip, overlapping images were
combined into a mosaic using the LINMOS routine in the MIRIAD
software package.  This routine does a simple linear mosaicing.  In
the four mosaic images, within the FWHM of the overlapping primary
beams, the rms noise is 42~$\mu$Jy/beam.  Initial source detection
was performed on these mosaic images using SExtractor with a
threshold of 143~$\mu$Jy/beam.  Each potential source was then
checked in the original images (not mosaics), and sources below
10$\sigma$ were removed from the list. This conservative detection
limit was needed to avoid spurious sources; many 5$\sigma$ bumps are
real, but many are sidelobes caused by the limited UV coverage.  Many
radio surveys with better sensitivity are available, including in the
EGS \citep{Ivison2006}, and the purpose of this paper is best
served by maximizing reliability.  We
used the AIPS task JMFIT to determine final sources positions and
flux densities and corrected the latter for delay beam distortions
\citep{Condon1998}.

\section{Radio Source Counterparts}

Altogether 51 radio components (some of which may be radio doubles)
were detected in an area of 0.5735~deg$^2$ (the 50\% primary beam
limit). This agrees well with the number of sources expected on the
basis of previous surveys (45, based on an average sensitivity limit
of 0.55~mJy/beam --- \citealt{Ciliegi2003}).  Table~1 lists the source
catalog and parameters of each source: name,\footnote
  {Radio sources from this paper are named EGS06 followed by the
  cardinal number given in Table~1.  The `06' in the name refers to the
  observation wavelength in centimeters.}
position, primary beam correction,
corrected total flux density, and deconvolved angular size if
the source is resolved.  For source pairs, components are listed separately
regardless of the likelihood of physical association.  Table~1
includes 37 isolated sources and seven pairs with separations in the
range 3\arcsec\ to 13\arcsec.  For three cases (10/11, 17/18, and
50/51), the radio morphology is that of a classical double.  For the
other five cases, we cannot tell from the radio morphology alone
whether they are classical doubles or separate sources.  For the
closest pairs, where the two components might be blended at the VLA
resolution, combined
flux densities are given as well as separate ones for each component.

\citet[hereafter FWKK]{FWKK} surveyed a small part of the present EGS
field to much deeper flux density, a completeness limit of 25~$\mu$Jy
at 5~GHz.  They found 8 sources within our coverage area and with
flux densities greater than 0.2~mJy.  Table~2 compares our results
with those of FWKK.  Our survey detected the four sources with the
highest flux densities but not the four faintest, as expected.  For
the seven pointlike sources, flux densities or upper limits agree
within the uncertainties except perhaps for EGS06~23=15V70, which might
have varied in the 14$+$ years between surveys.
Source~21=15V10 is extended, and it is hard to make a direct comparison
between the two surveys because they used different VLA
configurations.  All in all, the two surveys agree within their
limited area of overlap.

The position of each radio source was examined in the IRAC 3.6 to
8~\micron\ images of the EGS.  These images have angular resolution
(FWHM) of 2\arcsec\ and 5$\sigma$ sensitivities 0.9 to 6.2~$\mu$Jy
\citep{Barmby2006,Davis2006}.
Eight radio sources or pairs are outside the IRAC
coverage area.\footnote{
  Some of the sources ``outside the IRAC area'' have a few IRAC
  frames covering the position, but data are noisy and affected by
  cosmic rays.  Sources~5/6  show an IRAC source at the position of the
  northern radio component (6) but nothing close to the southern one (5)
  (which may be slightly extended or double in the radio). This could
  be a double radio source with very different lobe distances from
  the host galaxy, or the radio sources could be unrelated with only
  the northern one having an IRAC counterpart, or the southern radio
  source, which has low signal-to-noise, might not be real.  Most
  likely is that the IRAC observations are not deep enough to detect
  the true counterpart; despite that, we show the data for the source
  near EGS06~6 in tables and figures.  Sources~17/18 have a faint IRAC source
  at 3.6 and 4.5~\micron\ (2.7 and 3.7~$\mu$Jy respectively) 0\farcs9
  north of the northern source (18), but we do not consider this a
  detection of a counterpart.  Because the radio morphology is that of a
  classical double, we would expect the counterpart to be between
  the radio lobes.  Source~32 and 48 have IRAC sources nearby but too
  far from the radio positions to be deemed  counterparts.
  Deeper IRAC data would be needed to say more about these sources.}
Of 31 isolated radio sources inside the IRAC coverage area
(first part of Table~1), 28 have IRAC
counterparts within 0\farcs8, and the other three have no IRAC
counterpart closer than 1\farcs2.  We consider the first group
identified and the second group not.  Trials with random positions
suggest about 1.5\% chance of an IRAC source within 0\farcs8 of a
given position; i.e., zero to one of the IRAC counterparts may
be spurious on this basis. However, most of the proposed counterparts
have flux densities well above
the IRAC survey limit, and spurious matches involving
such bright sources are far less likely.  Of five possible source
pairs with IRAC coverage (second part of Table~1), the two pairs
with radio morphology of classical doubles both have an IRAC source
between the two radio components, as expected.\footnote{
  IRAC exposure time for 50/51 is only 1/10 normal, but the source
  has high flux density and was
   reliably detected.  However cosmic rays badly affect the
  5.8~\micron\ photometry.}
Thus for 33 ``easy'' cases, we find IRAC counterparts for 30, an
identification rate of 91\%.

\begin{figure}
\plotone{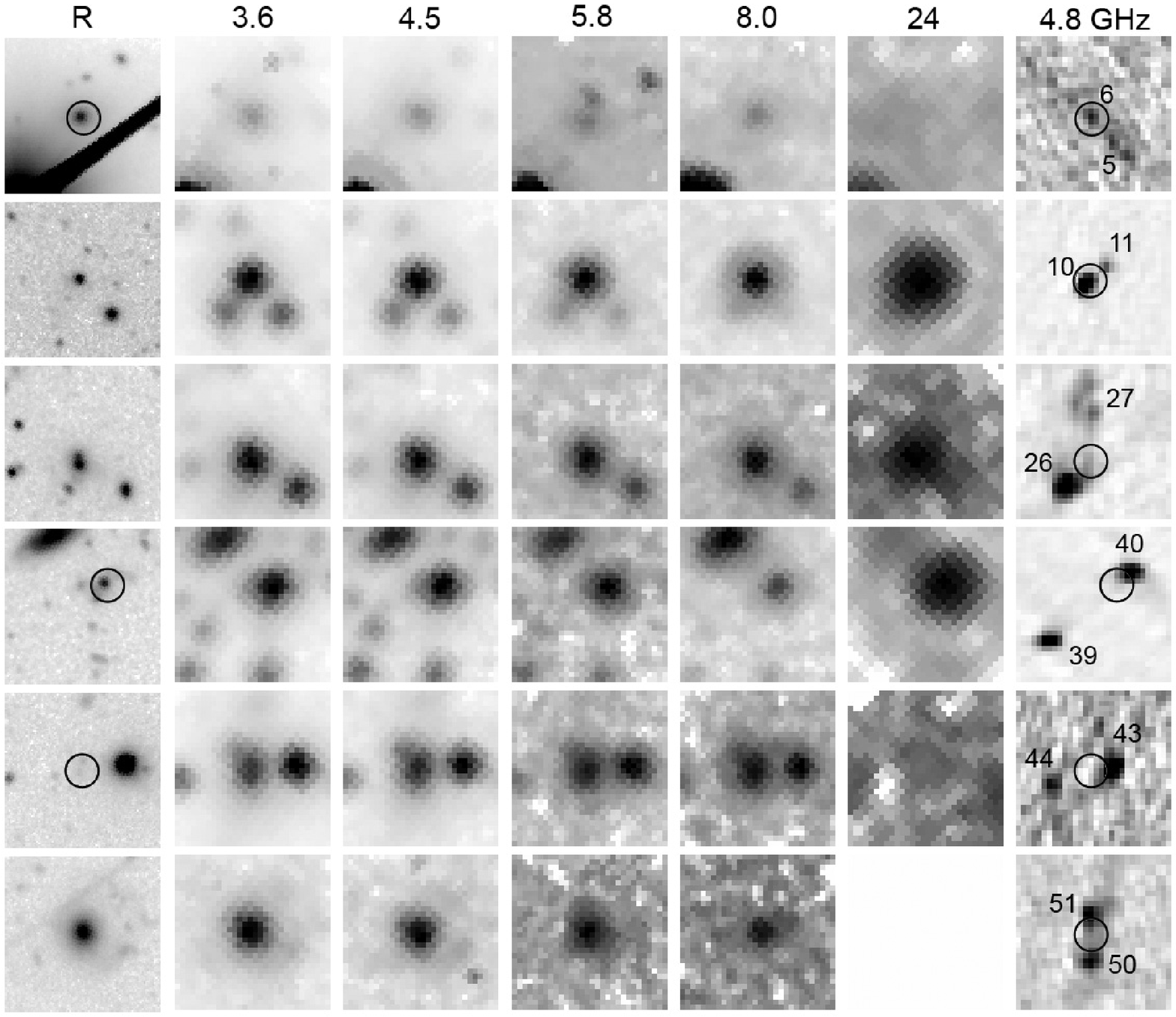}
\caption{Thumbnail images of possible double radio sources.
  Shown left to 
  right are Subaru $r$, IRAC 3.6, 4.5, 5.8, and 8.0~\micron, MIPS
  24~\micron, and VLA 4.8~GHz.  Each thumbnail is 18\arcsec\ on a
  side with north up and east to the left.  Greyscale is negative
  square root except for the radio images, which are negative
  linear. Circles show the location of 5\farcs2 photometry
  apertures. EGS06~50/51 are outside the MIPS 24~\micron\ coverage. }
\label{f:doubles}
\end{figure}

The three additional radio source pairs are more difficult to assess.
Source~26 is pointlike, but the 
nearby northern radio source (\#27) shows multiple sub-components.  An
IRAC source is 
4\farcs1 northeast of the southern radio source.  (See
Figure~\ref{f:doubles}.)  We consider this a
valid counterpart to a likely radio double, despite the IRAC galaxy
not being centered between the two radio sources. Radio sources
39/40 have wide separation, and there are additional, faint
IRAC sources in the field.  Any interpretation is uncertain, but we
deem the brightest IRAC source, located between the two radio
sources, a valid counterpart to a radio double.  There is an IRAC
source between sources~42/43, 2\farcs6 from \#43.  On higher resolution
red images from Subaru, this source is double with separation
1\farcs4.  We consider this a likely counterpart to the 42/43 pair.
Thus our best guess is that we have three IRAC counterparts to three
radio doubles and an overall identification rate of $33/36=92\%$.
While these three identifications are uncertain, they won't change
subsequent conclusions of this paper.  IRAC positions are listed in
Table~3.

The high identification rate ($>$90\%) in the infrared contrasts with
a much lower rate that would be found in visible light.  Only 24
radio sources have $R$ counterparts in the DEEP2 catalog ($R<23.95$,
\citealt{Davis2003}; see also \citealt{Coil2004} and
http://deep.berkeley.edu/), and of
these, only 12 are brighter than $R=21.5$.  The faintness of the
counterparts to 5~GHz radio sources found here contrasts with the
results of a 1.4~GHz survey \citep{Mobasher1999}, where half the
radio sources were found to have visible counterparts with $R<21.5$.
A more direct but smaller comparison sample is a 5~GHz survey in the
Lockman Hole \citep{Ciliegi2003}.  In that survey, of 12 radio
sources with $S_{5~GHz} > 0.2$~mJy, 8 have counterparts with $I<21$,
although \citeauthor{Ciliegi2003} accepted larger position offsets
than we do.  In contrast, we find counterparts that bright in $I$ for
only 14 of the EGS radio sources (with $S_{5~GHz} > 0.4$~mJy).
Considering that the radio surveys have different depths, different
wavelengths, different counterpart identification criteria, and are
done in different regions of sky, there are probably no major
inconsistencies in the fraction of counterparts optically identified.

Photometry of each IRAC counterpart was done on the four IRAC images
and the MIPS 24~\micron\ image.\footnote{Sensitivity at 70 and
  160~\micron\ was not sufficient to say anything useful about
  individual sources.}  (The MIPS images are also described
by \citealt{Barmby2006}.)  A center position was measured on
whichever image offered the best signal to noise ratio (usually
3.6~\micron\ but sometimes longer wavelengths), and the IRAF task
{\sc apphot} was used to measure flux in a 5\farcs2 diameter beam.
Sky was measured in an annulus of radius between 17$''$ and 24$''$,
though the exact choice did not matter.  The IRAC sources are all
pointlike, and flux densities are given in Table~4 based on the point
source calibration.  Except for sources near the edge of coverage
(indicated in Table~4), all the detected counterparts are bright
relative to the limiting magnitude of the images, and statistical
uncertainties (given in the last row of Table~4) are small.
Systematic uncertainties, due for example 
to centering errors or companion sources, are typically $\ll$5\% but can
reach 10\% for objects with nearby companions.  Calibration
uncertainty is about 3\% \citep{Reach2005}.

\begin{figure}
\plotone{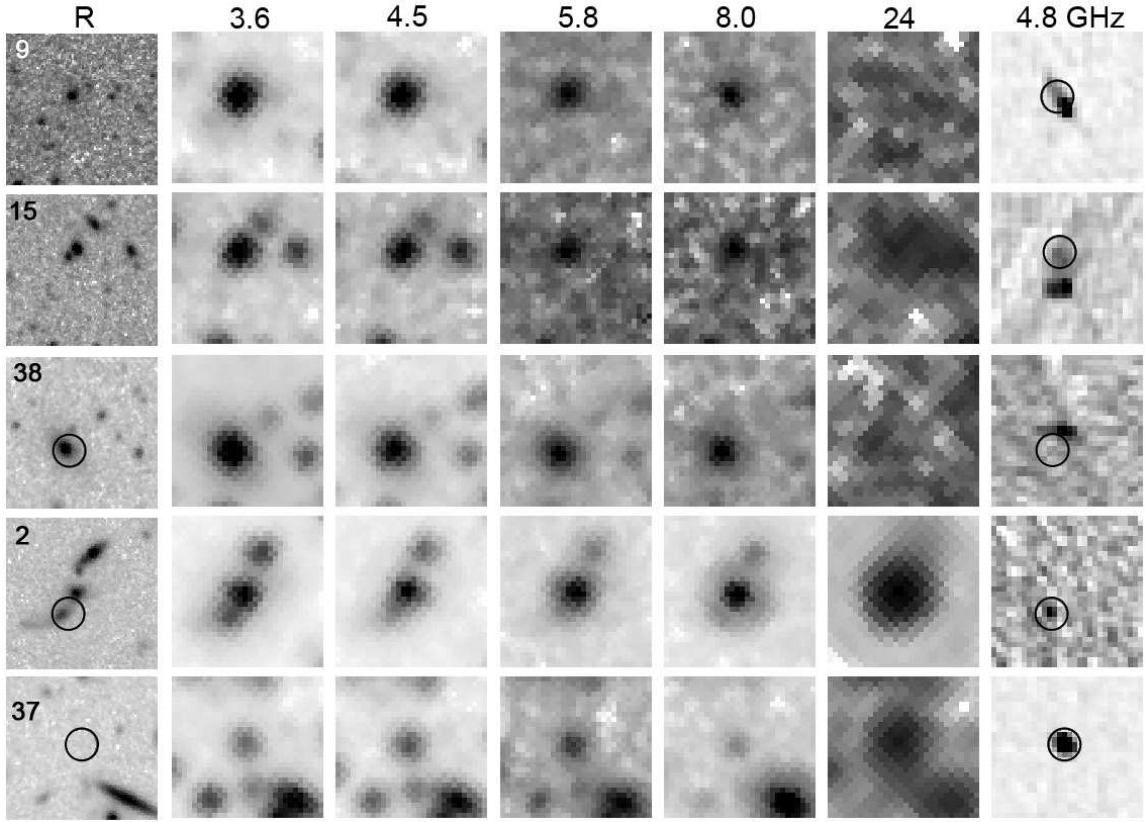}
\caption{Thumbnail images of radio source areas. Image details are
  the same as Fig.~\ref{f:doubles}. The top three rows show the
  regions of three radio sources considered unidentified.  Bottom
  two rows show interesting identified sources.  Circles show the
  location of the 5\farcs2 photometry apertures, the nearest IRAC
  source in the case of the three unidentified radio sources.
  The image of EGS06~2 is centered on the brightest IRAC component at
  14:14:33.16  +52:02:55.3 (J2000). }
\label{f:unid}
\end{figure}

Within the well-observed IRAC region, there are only three radio
sources that may not have IRAC counterparts. Fig.~\ref{f:unid} shows
thumbnails of these. Sources~9 and 15 show faint radio extensions to
the northeast and north, respectively.  In each case, there is an
IRAC and MIPS source within the radio extension 1\farcs8 and 4\farcs1
respectively from the radio peak positions.\footnote{
  Source~9 has full-depth IRAC coverage. Upper limits for a counterpart
  coincident with the radio peak are 0.9~$\mu$Jy at 3.6 and
  4.5~\micron\ and 6~$\mu$Jy at 5.8 and 8~\micron, 5$\sigma$.  Source~15
  is closer to the edge of the IRAC coverage; exposure time is about
  1/3 normal at 3.6 and 5.8~\micron\ and about 60\% of normal at 4.5
  and 8.0~\micron\ with detection limits correspondingly higher.}
While these sources could be the radio counterparts, establishing
them as such would require more evidence.  Source~38 shows a bright
galaxy (167~$\mu$Jy at 3.6~\micron) located 2\farcs6 southeast of the
radio position.\footnote{
  This position has full-depth IRAC coverage.}
The high-resolution $r$ image shows a much fainter galaxy ($r > 24$)
1\farcs4 northwest of the bright one, 1\farcs2 east of the radio
source peak and within an eastward radio extension.  This faint
galaxy could  be the counterpart, but at the low angular
resolution of the infrared images it
is difficult to distinguish from the bright galaxy
nearby.  The bright galaxy itself is extremely red, having $r-[3.6]
\approx 5.7$.  It could perhaps be the radio counterpart, but the
large offset from the radio position would be difficult to explain.

If there are radio sources without IRAC counterparts, they
are unlikely to be local galaxies \citep[cf.,][]{Masci2001}.
\citet{Gruppioni2001} suggested that radio sources with no
counterparts in deep optical surveys are likely to be early-type
galaxies at $z>1$.  However, the absence of an IRAC counterpart for
these galaxies is difficult to explain.  At $z=1$, the IRAC
3.6~\micron\ detection limit 0.7~$\mu$Jy corresponds to a stellar
luminosity of order $10^9$~\Lsol, and such low luminosity galaxies
are unlikely to harbor powerful radio sources.  Even at $z=5$, a
galaxy with $L=10^{11}$~\Lsol\ should have been detected at
4.5~\micron.  Any radio sources without IRAC counterparts are thus
either less luminous or more distant than these limits.

\section{Discussion}

\begin{figure}
\plotone{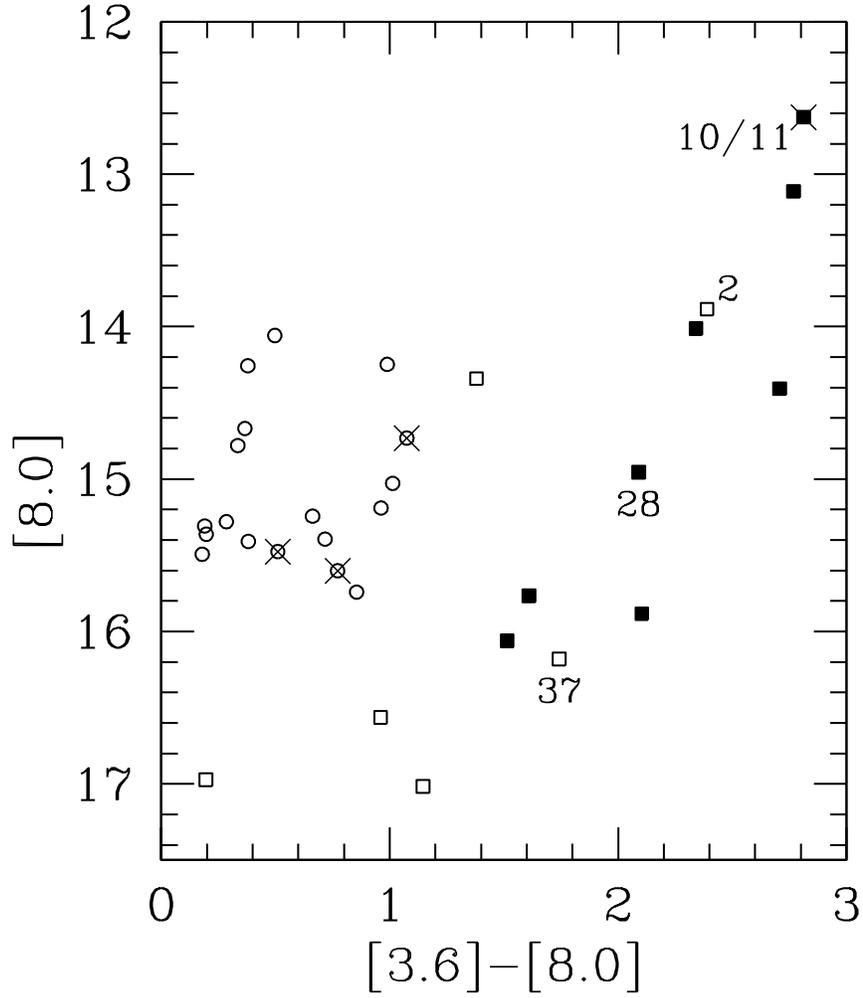}
\caption{Color-magnitude diagram for IRAC counterparts of radio
  sources.  Filled symbols indicate galaxies classified as AGN
  because they have power-law SEDs.  Symbols with superposed x
  indicate the four detected X-ray sources.  The ``blue
  clump'' refers to the group of sources with $[8.0]<16$ and
  $[3.6]-[8.0]<1.2$, indicated with open circles. Four objects
  discussed in the text are identified with numbers from Table~3.
  The 5$\sigma$ sensitivity limit is $[8.0]= 17.55$; sensitivity at
  3.6~\micron\ does not limit detection of sources plotted here.}
\label{f:cm4}
\end{figure}

Spectral energy distributions (SEDs) indicate the type of galaxy and
likely redshift range.  Apparent magnitude is also a good distance
indicator for radio galaxies \citep[e.g., Fig.~1 of][]{vanB1999}.
Fig.~\ref{f:cm4} shows an IRAC color-magnitude diagram for the radio
source counterparts.  There are two distinct groups of galaxies.  One
group, the ``blue clump,'' is brighter than $[8.0]=15.8$ and quite
blue ($[3.6]-[8.0]<1.1)$.  A second group follows a distinct
``stripe'' from faint and blue to bright and red.  Similar, though
less distinct, separations are present in other C-M diagrams.

\begin{figure}
\plotone{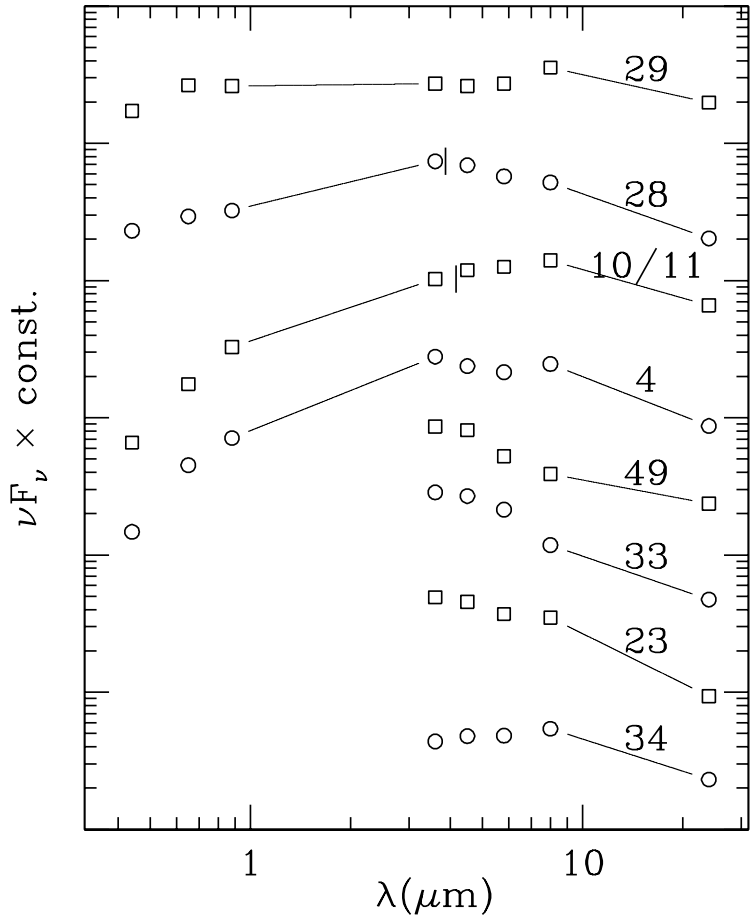}
\caption{SEDs for eight galaxies showing power law SEDs and thus
  classified AGN.  Wavelengths are in the observed frame;
  vertical lines show rest 1.6~\micron\ for galaxies
  having spectroscopic redshifts.  Symbol types alternate for
  clarity. Lines bridge wavelength gaps in the observations and
  should not be taken to suggest the actual SED.  Photometric error
  bars are smaller than the symbol sizes in this and subsequent SED plots.}
\label{f:agn}
\end{figure}

Eight of the radio source counterparts have distinct power law SEDs
(Fig.~\ref{f:agn}): nearly a straight line in the log-log plot.
(\citealt{Davis2006} describe the BRI data used in the SED plots.)  
All these sources have spectral index $\alpha < 0$ ($F_\nu \propto
\nu^\alpha$).  Such SEDs are signatures of powerful AGN
or QSOs \citep[e.g.][]{Elvis1994}. Two of these galaxies are
confirmed spectroscopically as broad-line AGN (Table~3); there are no
spectra for the other six.  All eight of the power-law galaxies are
red in the [3.6]$-$[8.0] color and fall along the ``stripe'' in the
C-M diagram with the brightest AGN being the reddest.  This may
represent the combined effects of AGN luminosity and galaxy distance.
Luminous, nearby AGN have little contribution from starlight at
8~\micron\ and are therefore bright and red.  Less luminous AGN are
fainter, and the greater contribution of starlight relative to the
AGN makes the color bluer.  More distant AGN are also fainter, but
because the observed radiation emerged at shorter wavelength in the galaxy's
rest frame, starlight makes a bigger contribution than for nearby
galaxies.

\begin{figure}
\plotone{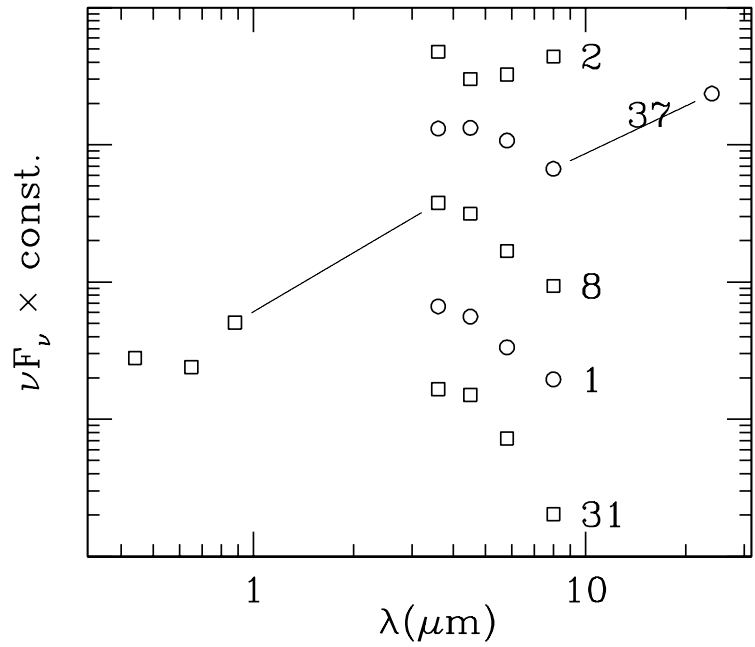}
\caption{SEDs for five galaxies in the stripe of the C-M diagram but
  not having power law SEDs.  Wavelengths are in the observed
  frame; none of these galaxies has a
  spectroscopic redshift. For EGS06~2, the data shown are for only
  the single component shown in Fig.~\ref{f:unid}, not for the entire
  system.  Symbol types alternate for clarity. Lines bridge
  wavelength gaps in the observations and should not be taken to
  suggest the actual SED.}
\label{f:stripe}
\end{figure}

Two other galaxies occupy the stripe region of the C-M diagram along with  the AGN.
One of them (\#2) is a component of a dusty, interacting system.
The infrared component corresponding to the radio source has a near
power-law SED (Fig.~\ref{f:stripe}) but with an upturn at
3.6~\micron.  This component could be an AGN with contribution from
starlight at the shortest wavelength.  Fig.~\ref{f:unid} shows
thumbnail images of this source.

The other source in the stripe, \#37, is unusual.  It is easily
detected in all IRAC bands with a flat SED (Fig.~\ref{f:stripe}), but
it is invisible on I-band HST images ($I>26.8$, 0.04$\mu$Jy).  The
source is bright at 24~\micron\ and has the second-highest radio flux
density in the survey.  One possibility is that \#37 is a dusty
galaxy at $2\la z\la 3$, consistent with the $[3.6]-[4.5]$ color and
the 1.6~\micron\ stellar emission peak being observed near 5~\micron\
(Fig.~\ref{f:stripe}). The observed 24~\micron\ would then come from
the PAH emission features at rest 6--8~\micron, and the faint V and I
magnitudes could be attributed to a combination of the Balmer
spectral break and heavy dust reddening.  \citet{Higdon2005} reported
finding four radio sources with similar 24~\micron\ flux densities
and upper limits in visible light, but \#37 has much higher radio
flux density than any of the Higdon et~al.\ sources.
\citet{Houck2005} gave redshifts for 17 $R>24.5$ MIPS
sources, but only one would have been detectable in our radio survey.
That source (their \#13 at $z=1.95$) is ten times brighter at 24~\micron\ but
perhaps eight times fainter at 6~cm (guessing a radio spectral index
of $-0.7$) than our \#37.  An
alternative possibility is that \#37 could be similar to extreme
radio galaxies reported by \citet{vanB1999} and
\citet{Waddington1999}.  If the intrinsic $K-[3.6]$ color is not extreme and
the usual magnitude-redshift relation \citep{vanB1999} applies,
the magnitude $[3.6]=17.9$ suggests
$z\approx2$ but is consistent with redshift as large as 5.

\begin{figure}
\plotone{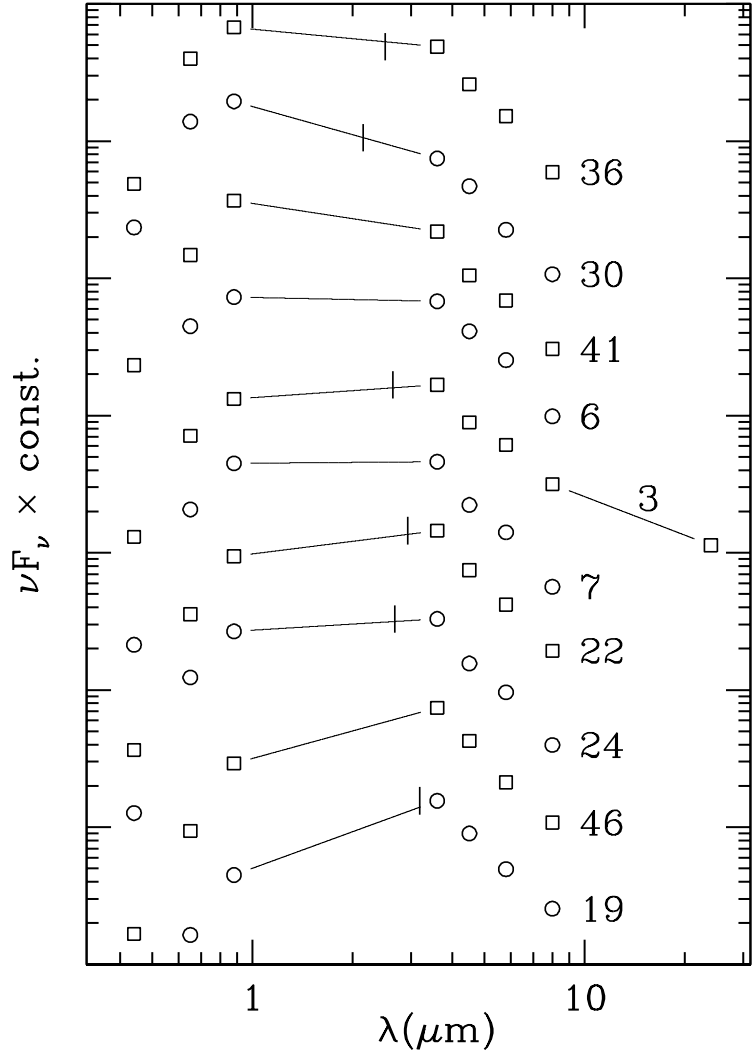}
\caption{SEDs for ten 
  blue clump galaxies having colors most resembling the
  elliptical template in Fig.~\ref{f:cc1}. Symbol types alternate for
  clarity.  Wavelengths are in the observed frame; vertical lines
  show rest 1.6~\micron\ for galaxies having 
  spectroscopic redshifts.  Symbol types alternate for clarity. Lines
  bridge wavelength gaps in the observations and should not be taken
  to suggest the actual SED.}
\label{f:clump1}
\end{figure}

\begin{figure}
\plotone{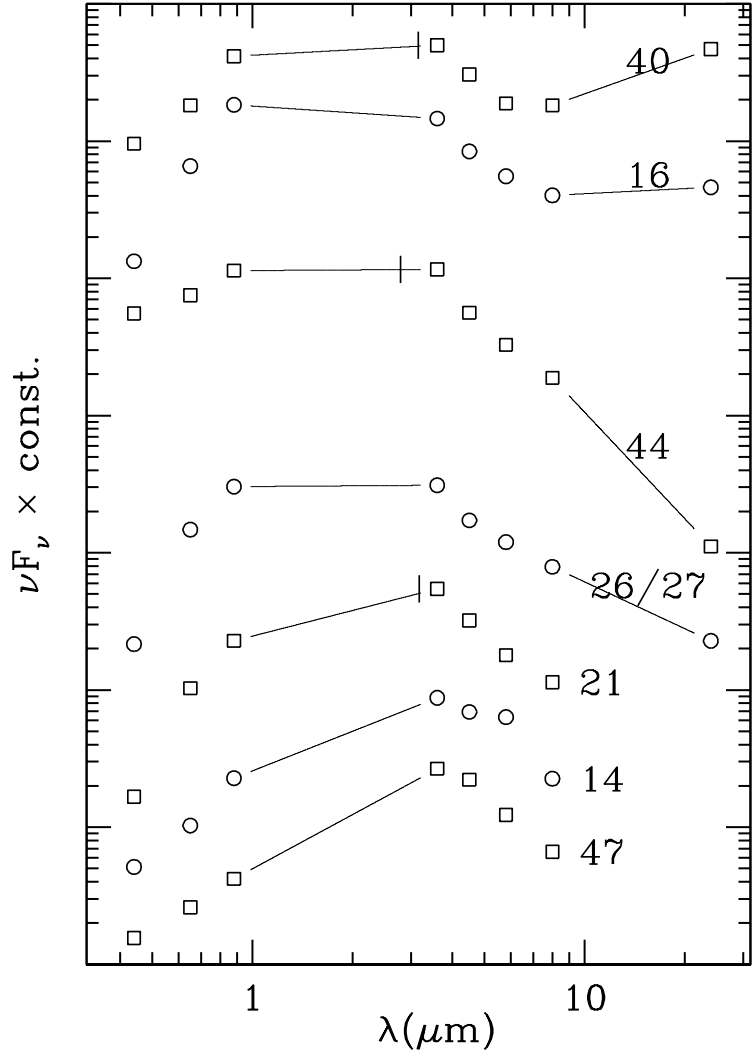}
\caption{SEDs for six 
  blue clump galaxies having colors most resembling the
  spiral template in Fig.~\ref{f:cc1}. Symbol types alternate for
  clarity.  Wavelengths are in the observed frame; vertical lines
  show rest 1.6~\micron\ for galaxies 
  having spectroscopic redshifts.}
\label{f:clump3}
\end{figure}

\begin{figure}
\plotone{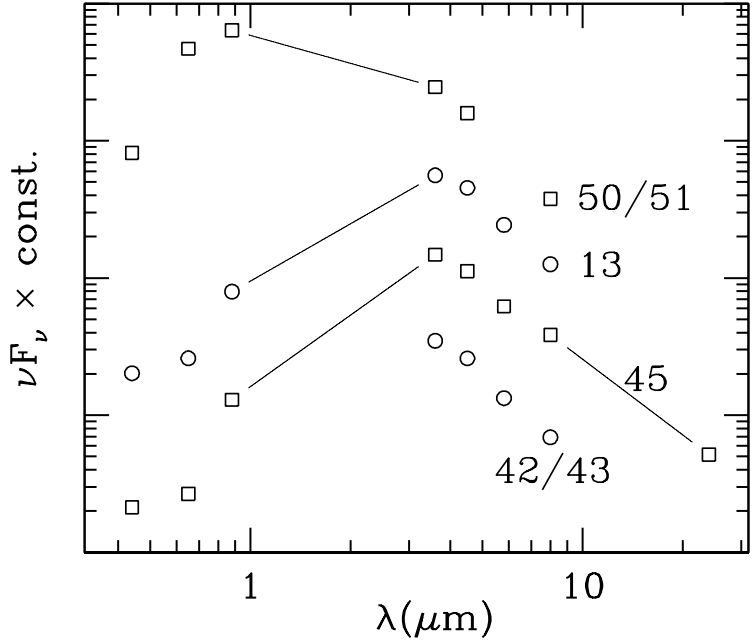}
\caption{SEDs for four 
  blue clump having colors that cannot
  distinguish a likely morphological
  type.  Red $[3.6]-[4.5]$ colors suggest $z>1$ for all.
  Symbol types alternate for clarity. Lines bridge wavelength gaps in
  the observations and should not be taken to suggest the actual
  SED.  Wavelengths are in the observed frame; none of these galaxies
  has a spectroscopic redshift.}
\label{f:clump2}
\end{figure}

\begin{figure}
\includegraphics[width=6.5in,trim= 72 216 0 0]{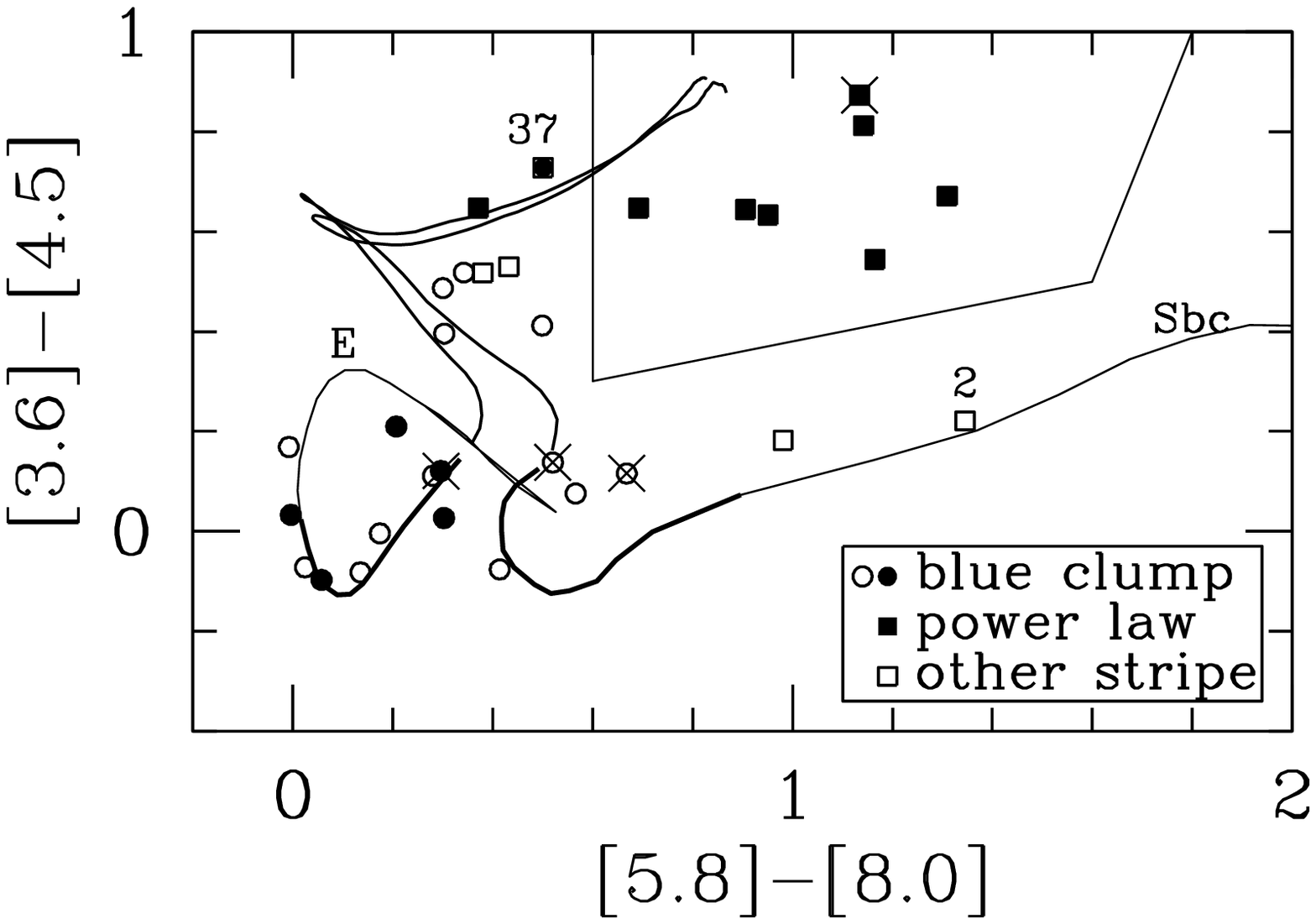}
\caption{Color-color diagram for IRAC counterparts of radio sources.
  Circles indicate galaxies in the
  blue clump; filled circles indicate blue clump galaxies
  with spectra that show an old stellar population.
  Filled squares indicate galaxies classified as AGN because they
  have power-law SEDs.  Open squares indicate the remaining galaxies
  in the stripe; they have color/magnitude combinations outside the
  blue clump but not power law SEDs. X-ray sources are marked with a
  superposed x. Curved lines show colors of typical Sbc and E
  galaxies as a function of redshift from $z=0$ to $z=4$; curves are
  labelled near the low-redshift end.  The heavy portion of each
  track indicates $0.5 < z < 1$, and $z=4$ is at the top of the
  diagram.  The color tracks show PAH emission shifting out of the
  IRAC 8~\micron\ band as redshift increases from 0 to 0.5 and the
  1.6~\micron\ peak of the stellar energy distribution shifting
  through the IRAC bands as redshift increases.  Thin, straight lines
  indicate the area \citet{Stern2005} found for the colors of
  spectroscopically-identified AGN. Objects with uncertain photometry
  as indicated in Table~4 are omitted, and two objects of interest
  are labeled.}
\label{f:cc1}
\end{figure}

\begin{figure}
\includegraphics[width=6.5in,trim= 54 144 0 0]{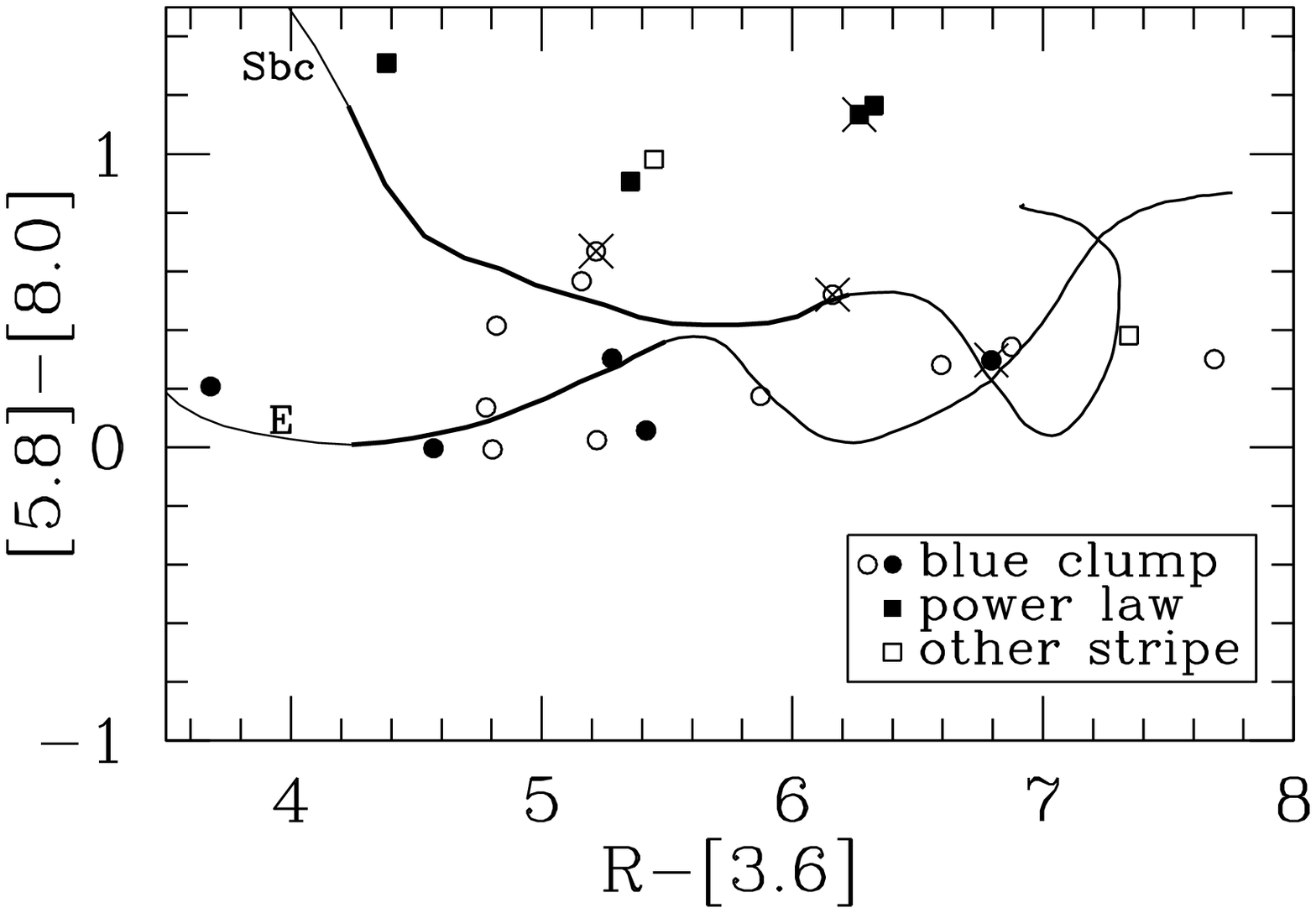}
\caption{Color-color diagram for IRAC counterparts of radio sources.
  Circles indicate galaxies in the
  blue clump, and filled circles indicate blue clump galaxies
  with spectra that show an old stellar population.
  Filled squares indicate galaxies classified as AGN because they
  have power-law SEDs.  Open squares indicate the remaining
  galaxies, which have color/magnitude combinations outside the blue
  clump but not power law SEDs.  X-ray sources are marked with a
  superposed x.  Curved lines show colors of Sbc (red
  $[5.8]-[8.0]$ color) and E galaxies as a function of redshift from
  low redshift to $z=4$.  The heavy portion of each track indicates
  $0.5 < z < 1$, and tracks are labelled near $z=0.5$. Objects
  with uncertain photometry as indicated in Table~4 are omitted.  The
  red colors of many of the counterparts suggest high redshifts and
  underscore the difficulty of detecting counterparts in
  visible-wavelength images.  The object with $R-[3.6]<4$ is \#30 at
  $z=0.346$.}
\label{f:cc6}
\end{figure}

The galaxies in the ``blue clump'' in the color-magnitude diagram are
all well above the sensitivity limit of the IRAC data.
\citet{Waddington2000} identified counterparts of radio galaxies
found at 1.4~GHz and found a peak in 2.2~\micron\ number counts at
$K\approx16$.  This peak almost certainly represents the same
population as the blue clump, but \citeauthor{Waddington2000} could
say little about these galaxies without data at longer wavelengths.
The blue clump galaxies have SEDs consistent with normal galaxies
(Figs.~\ref{f:clump1}--\ref{f:clump2}).  The color-color diagrams in
Figs.~\ref{f:cc1} and~\ref{f:cc6} compare these galaxies to the
colors of typical spiral and elliptical galaxies.  Most of the
galaxies have $0.5<z<1$ as indicated by the color tracks in
Figs.~\ref{f:cc1} and~\ref{f:cc6}, primarily the $[3.6]-[4.5]$ color.
The 3.6~\micron\ magnitudes of the blue clump galaxies range from
14.5 to 16.5, consistent with the redshift-K relation
\citep{vanB1999} for $0.5<z<1$ and the expected $K-[3.6] \approx 0.5$
at these redshifts.  If $z>0.5$, and assuming a radio spectral index not
far from the usual $-0.7$, the radio power emitted at 1.4~GHz is
$>$10$^{24}$~W~Hz$^{-1}$.  Such high radio luminosity indicates an
AGN, consistent with the findings of \citet{Hammer1995} and
\citet{Mobasher1999} in their radio surveys.  \citet{Benn1993} found
a much larger fraction of star-forming galaxies, but their data apply
to fainter radio flux densities and brighter optical magnitudes,
hence lower redshifts, than the sources studied here.  The absence
of an obvious power law emission component in the mid-infrared
implies that the active nucleus is either obscured or has such low luminosity
that it does not dominate the near infrared SED of the galaxy.
\citet{Barmby2006} found that 60\% of X-ray selected AGN also lack a
dominant mid-IR power law component.  No doubt similar galaxies exist
at larger redshifts than are detected here, but they are below the
detection limit of the radio survey because of the strong
K-correction as well as inverse-square dimming.

\begin{figure}
\plotone{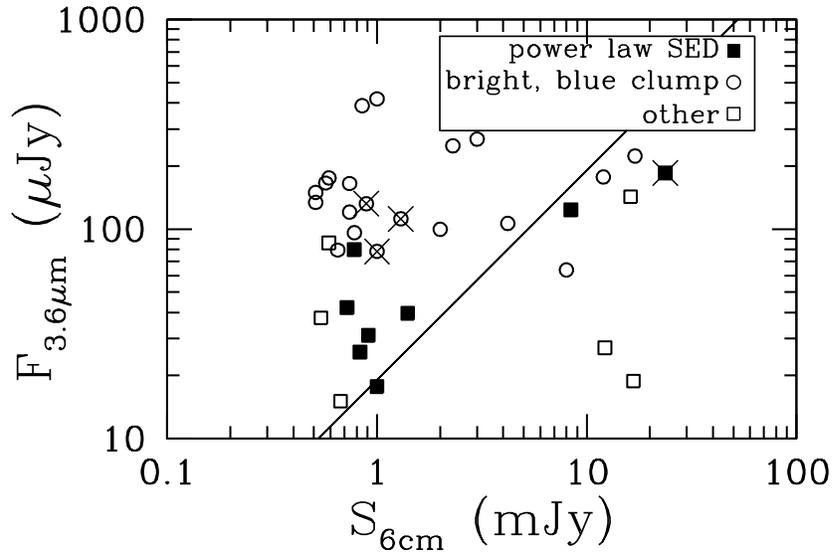}
\caption{Comparison of 3.6~\micron\ and radio flux densities.  Filled
  symbols indicate galaxies classified as AGN because they have
  power-law SEDs.  Open circles indicate galaxies in the blue clump.
  Open squares indicate the remaining galaxies, which have
  color/magnitude combinations outside the blue clump but not power
  law SEDs.  X-ray sources are marked with a superposed x.  The solid
  line shows the average ratio of 3.6~\micron\ to radio flux density
  for radio loud quasars \citep[Fig.~10]{Elvis1994}.  
  Galaxies undetected at 3.6~\micron\ would
  fall more than an order of magnitude below the bottom
  limit of this plot.}
\label{f:rad_irac1}
\end{figure}

The galaxy classes separated in the color-magnitude diagram also show
different ratios of radio to near-infrared flux density.
Fig.~\ref{f:rad_irac1} shows that for a given radio flux density, the
blue clump galaxies tend to be brighter at 3.6~\micron\ than the
stripe galaxies.  \citet[their Fig.~12]{Georgakakis1999} found that
the  radio sources with the highest flux densities
(which are the only ones our survey could
have detected) tended to be AGN and absorption systems (i.e.,
ellipticals) rather than star-forming galaxies.  Presumably the radio
emission comes from an AGN in these types.  \citet{Masci2001} also
found radio emission larger than expected from star formation alone
in their radio sample. If radio flux density indicates ``AGN power,''
Fig.~\ref{f:rad_irac1} suggests that the clump galaxies contain a
large stellar population that outshines the underlying AGN.  Further
observations (and in particular redshifts) will be needed to
determine whether the ``blue clump'' and ``stripe'' populations differ in
fundamental ways or are simply the result of varying 
ratios of AGN to stellar light.

Photometric classification of galaxies is an uncertain business at
best, but Figs.~\ref{f:cc1} and~\ref{f:cc6} show there is a tendency
for more blue-clump galaxies to fall close to the elliptical than the
spiral colors.  The colors also provide very rough photometric
redshifts.  Figs.~\ref{f:clump1} and~\ref{f:clump3} illustrate the
SEDs.  In all cases where DEEP2 spectroscopy is available, it confirms the
photometric classifications.\footnote{Most of the 
  spectra were not yet available when the photometric classifications
  were made, so this is not a case of ``knowing the answer in advance.''}
The spectra exhibit three main classes:
\begin{enumerate}
\item
An old stellar population (labelled ``old'' in Table~3). These
galaxies show calcium H and K lines in absorption and often the
G~band, consistent with high luminosity, early-type galaxies.  A few
of these galaxies show weak [\ion{O}{2}] emission lines indicative of
some recent star formation or LINER activity, but the $U-B$ colors
are red ($\ge$0.33), 
suggesting that the star formation or nuclear activity has had little
effect on the 
bulk of the stellar population.  Of the ten galaxies in
Fig.~\ref{f:clump1}, the six that  have spectra are all of old type.
\item
Three galaxies show strong Balmer absorption lines  typical of
A~stars; these galaxies are 
labelled ``post-SB'' (post-starburst) in Table~3.  All three are in the
blue clump and have colors resembling the spiral template.  (See
Fig.~\ref{f:clump3}.)  One of the three galaxies (\#39/40) also shows
[\ion{Ne}{5}] emission lines characteristic of AGN.  
Its SED shows a rise at 8~\micron,
which may be the signature of a power-law spectral component.
\item
Three galaxies show AGN emission lines.  Two with broad emission lines 
show power law SEDs (Fig.~\ref{f:agn}), and the third is \#39/40
mentioned above.
\end{enumerate}
Appendix~A gives details of all the spectra.  The apparent numbers of
ellipticals and spirals are consistent with previous results
\citep{Georgakakis1999, Gruppioni1999, Ciliegi2003, Richards1998,
Chapman2003}.  The prevalence of early-type, i.e., not star forming,
galaxies adds weight to our suggestion that  the radio sources
arise from active nuclei, not starbursts.  Despite the good agreement
between photometry and 
spectra as regards galaxy type, there is one discrepancy in the
redshift order. Source~3 has a higher redshift than \#22 according to
color, but spectroscopy shows the opposite.  This should serve as a
warning to be cautious about classifications or redshifts derived
from limited photometry.  Nevertheless, the spectroscopic redshifts
confirm the general redshift range $0.5 \la z \la 1.0$ derived from
colors and magnitudes.

Another indicator of nuclear activity is X-ray emission
\citep{mush04}.  Published X-ray observations with Chandra
\citep{egs_cxo} and XMM-Newton \citep{was03} cover part of the EGS
\citep{Barmby2006}.  Six radio sources are within the area covered by
both Chandra and XMM: four (EGS06~14/22/23/24) are not
detected in X-rays. Source~19 is coincident with Chandra source 83 but is
not detected by XMM; source~21 coincides with Chandra source 55 and XMM
source 53. Eight radio sources (eleven components) are within the area
covered by XMM alone: six (EGS06~12/13/15/17/18/20/26/27) are not
detected in X-rays, but EGS06~10/11 and 16 are coincident with XMM
sources 5 and 43, respectively.  The X-ray/radio coincidence rate is
therefore 4/14, or 29\%.  EGS06~10/11 has the highest radio flux density in
the sample and is a definite AGN from its SED.  The other three
X-ray-detected sources are clump sources with no obvious distinction
from the other clump sources except that \#16 shows modest excess
emission at 8~\micron\ that might be attributed to an AGN power
law. The ratio of 24~\micron\ flux density to soft X-ray flux puts all
four X-ray sources into the AGN category by an order of magnitude
(\citealt{Weedman2004}; $IR/X<0.05$ in their units).
EGS06~10/11 and 23 are the only radio 
sources in the X-ray area whose IRAC SEDs are AGN-like. A possible
explanation for the X-ray non-detection of 23 is obscuration, as
found by \citet{don05} for ``radio-excess AGN.''  Detection of radio
sources that are likely to be AGN yet do not show X-ray emission is 
additional evidence that no single-wavelength survey can produce a complete
list of AGN.

\section{Conclusions}

IRAC images are a powerful means of identifying and classifying radio
sources.  Deep IRAC images identify counterparts to at least 92\% and
possibly 100\% of $\sim$0.5~mJy radio sources.  Many of the
counterparts have very red visible to mid-infrared colors, 
underscoring the difficulty of
identification at shorter wavelengths.  The radio sources are 
likely to arise from powerful AGN, but only about 25\% show a typical
AGN power-law component that dominates the mid-infrared emission.
For about 70\% of the counterparts, galaxy starlight dominates the
mid-infrared emission.  The starlight-dominated galaxies can be
identified by their blue $[3.6]-[8.0]$ IRAC colors.  These ``blue
clump'' galaxies show a small range of apparent magnitude, suggesting
a relatively small range of redshift for this galaxy type in this
survey.  The small range is probably a combination of the radio
survey sensitivity limit and relative rarity at low
redshift of galaxies hosting 
powerful radio sources.  The separation into ``stripe'' and ``blue
clump'' galaxies 
should be useful in classifying galaxies found in other radio
surveys.




\acknowledgments

We thank Jeff Newman and the DEEP team for access to the Keck
spectroscopic data.  This work is based in part on
observations made with the Spitzer Space Telescope, which is operated
by the Jet Propulsion Laboratory, California Institute of Technology
under a contract with NASA. Support for this work was provided by
NASA through an award issued by JPL/Caltech.  The National Radio
Astronomy Observatory is a facility of the National Science
Foundation operated under cooperative agreement by Associated
Universities, Inc.  A.L.C. acknowledges support by NASA through
Hubble Fellowship grant HF-01182.01-A, awarded by the Space Telescope
Science Institute, which is operated by the Association of
Universities for Research in Astronomy, Inc., for NASA, under
contract NAS 5-26555.  DEEP spectroscopy is supported by the
National Science Foundation grants AST-0071198 and AST-0507483.



Facilities:
\facility{VLA}
\facility{Spitzer/IRAC}
\facility{Spitzer/MIPS}
\facility{Keck}
\facility{Subaru}



\appendix

\section{Details of DEEP2 Visible Spectra}

Spectra are all of good quality. Whenever the [\ion{O}{2}] doublet is
seen with good S/N, the lines have velocity widths too broad for a
clean separation of the doublet components, which are 220~km~s$^{-1}$
apart. The implied large internal kinematics are consistent with the
prevailing view that radio sources are hosted by a massive galaxies.

For each object that has a spectrum, the DEEP2 catalog number, mask,
and slit identifications are listed in parentheses followed by a
brief description of the spectrum. Objects are identified by their
EGS06 numbers as given  in Tables 1 and~3. 

3:
(11021233, mask 1101, slit 111)
Absorption spectrum characteristic of old population shows  Ca~HK,
H$\delta$, G~band, H$\beta$, and Mg~b.

19:
(12007962, mask 1245, slit 85)
The absorption line spectrum showing Ca~HK is consistent with an old
stellar population; moderate-strength 
[\ion{O}{2}] emission is also seen.

21:
(12012467, mask 1243,   slit 88)
Moderate strength [\ion{O}{2}] is seen along with moderately strong
Balmer absorption lines of H$\delta$ and H$\gamma$. Spectral type is
between an old and post-starburst stellar population.

22:
(12012898, mask~1205,	slit~107)
Excellent match to an old stellar population template through
H$\beta$ and [\ion{O}{3}]; [\ion{O}{2}] emission is visible but of
too low S/N to estimate a velocity width.

24:
(12016405, mask~1209,  slit~72)
Excellent match to an old stellar population template
from Ca~H through Mg~b; [\ion{O}{2}] is out of spectral range.

28:
(12020403, mask~1210,	slit~97)
AGN signatures are seen with \ion{Mg}{2}, [\ion{Ne}{5}], and
[\ion{O}{2}] all in emission and broad.  Two sets of strong
\ion{Mg}{2} absorption are superimposed with one that is close to the
redshift defined by the [\ion{O}{2}] emission line and another that
is shifted to the red (perhaps from infalling gas) by about
300~km~s$^{-1}$.  There is also a "foreground" \ion{Mg}{2} absorption
doublet seen at redshift z=1.38; candidates for its source may be
among galaxies seen in the field at separations of about 3\arcsec.

30:
(13004312, mask~1300,	slit~73)
Good match to a pure absorption-line   old stellar
population over the wavelength range  from near H$\beta$ through H$\alpha$. 

36:
(13025514, mask~1309,	slit~49)
Excellent match to a pure  old stellar population over the wavelength 
range from the G~band
through \ion{Na}{1}. Faint
continuum of another spectrum  $\sim$2\arcsec\ away shows a slight
absorption at the G~band. Being at the same redshift as that of the 
radio galaxy, this feature suggests that we may be watching
a minor merger.

44:
(13058191, mask~1315,	slit~121)
Excellent S/N continuum shows unusually strong Balmer absorption
indicative of a post-starburst. [\ion{O}{2}] is outside the spectral
range, and the red end is past Mg~b.  H$\beta$ and H$\gamma$ emission
lines are also seen; they appear tilted in the 2-D spectrum and are
spatially asymmetric. Moreover, the H$\beta$ emission is redshifted
by several 10's of km~s$^{-1}$ with respect to its absorption,
consistent with a possible inflow of ionized gas.  [\ion{O}{3}] lines
are not seen.  The r~image shows no companions within 3\arcsec.

10/11:
(11045619, mask~1114,	slit~83)
The spectrum has broad emission lines of \ion{Mg}{2} and
[\ion{Ne}{5}], as well as strong \ion{Mg}{2} and \ion{Mg}{1} in
absorption.  If the emission lines of [\ion{Ne}{5}] are used as a
reference, these absorption lines of \ion{Mg}{2} show a redshift by
150~km~s$^{-1}$, while a much weaker set of \ion{Mg}{2} absorption
lines appear with a blueshift of over 500km/s.

39/40:
(13032337, mask~1313,	slit~30)
The spectrum has excellent S/N and shows unusually strong Balmer
absorption lines that are characteristic of post-starbursts. Strong
[\ion{Ne}{5}] emission lines at 3345\AA\ and 3426\AA\ characteristic
of AGN along with strong [\ion{O}{2}] are also detected with velocity
dispersions of about 200~km~s$^{-1}$.  This AGN/starburst spectrum is
flanked by two other spectra a few arcsecs away and at slightly lower
redshifts by about 100~km~s$^{-1}$. One shows narrow (unresolved)
emission lines of [\ion{O}{2}], H$\delta$, and H$\gamma$, while the
other shows broader lines ($\sigma\approx200$~km~s$^{-1}$) of
[\ion{O}{2}] and [\ion{Ne}{3}].  Yet another flanking feature is weak
detection of [\ion{O}{2}] emission at redshift z=1.42, presumably
from a background (possibly lensed) galaxy.  The r~image shows at
least four other galaxies at separations $\ga$3\arcsec.







\clearpage

\clearpage

\clearpage

\clearpage

\clearpage

\clearpage

\clearpage

\clearpage

\clearpage

\clearpage

\clearpage

\clearpage

\begin{deluxetable}{rcccccl}
\tabletypesize{\footnotesize}
\tablecaption{EGS 6~cm Radio Sources}
\tablehead{
\colhead{Name} &  
\colhead{RA} &
\colhead{Dec} &
\colhead{PB} &
\colhead{Stot} &
\colhead{Size(\arcsec)} &
\multicolumn{1}{l}{Comments}\\
\colhead{EGS06} &
\multicolumn{2}{c}{J2000} &
\colhead{corr} & 
\colhead{mJy} &
\colhead{maj~min}
}
\startdata
\multicolumn{4}{l}{Isolated sources}\\
   1  &14 15 03.92 &52 01 06.5 &1.14 &$  0.67 \pm .09 $&           & \\
   2  &14 14 33.36 &52 02 53.1 &1.12 &$  0.59 \pm .07 $&		 & \\
   3  &14 14 36.75 &52 05 03.1 &1.02 &$  0.74 \pm .10 $&  0.6 $<$0.3 & \\
   4  &14 15 08.83 &52 06 42.2 &1.16 &$  0.78 \pm .12 $&		 & \\
   7  &14 16 15.07 &52 11 21.6 &1.00 &$  0.57 \pm 0.1 $&           &    \\
   8  &14 16 11.86 &52 12 04.8 &1.02 &$  12.2 \pm 0.1 $&  0.6 $<$0.2 &  \\
   9  &14 15 36.84 &52 14 09.7 &1.15 &$  10.9 \pm 1.5 $&  5.0 $<$1.0 & pa 40\degr\\
  12  &14 16 08.06 &52 24 58.5 &1.24 &$  3.2  \pm 0.3 $&  1.0 $<$0.5 & outside IRAC area \\
  13  &14 16 21.88 &52 25 03.3 &1.06 &$  8.0  \pm 0.2 $&  0.9 $<$0.3 & \\
  14  &14 17 59.29 &52 25 54.0 &1.39 &$  0.65 \pm 0.2 $&  0.6 $<$0.2 & \\
  15  &14 16 30.12 &52 27 01.1 &1.15 &$  9.0  \pm 2   $&  5.0 $<$1.0 & peak flux position, extended to north \\
  16  &14 16 27.95 &52 27 07.1 &1.21 &$  0.89 \pm .14 $&		 & \\
  19  &14 17 49.20 &52 28 03.1 &1.13 &$  1.3  \pm 0.2 $&  1.1 $<$0.3 & \\
  20  &14 16 23.30 &52 28 27.2 &1.41 &$  1.1  \pm 0.2 $&           & outside IRAC area \\
  21  &14 17 32.63 &52 32 03.1 &1.04 &$  1.0  \pm 0.2 $&  1.9 $<$0.8 & \\
  22  &14 17 13.59 &52 32 13.8 &1.07 &$  0.74 \pm .09 $&		 & \\
  23  &14 18 09.95 &52 33 00.2 &1.19 &$  1.0  \pm 0.2 $&  1.8 $<$0.5 &  \\
  24  &14 17 53.38 &52 35 39.3 &1.02 &$  0.51 \pm 0.07$&           & \\
  25  &14 17 04.19 &52 37 54.5 &1.81 &$  3.7  \pm 0.5 $&  1.3 $<$0.6 & outside IRAC area  \\
  28  &14 18 05.55 &52 40 32.3 &1.19 &$  0.72 \pm .19 $&           & \\
  29  &14 19 45.50 &52 46 48.1 &1.33 &$  8.4  \pm 0.2 $&  0.3 $<$0.2 & \\
  30  &14 19 10.40 &52 48 30.7 &1.14 &$  3.0  \pm 0.6 $&  7.4 1.5  & pa 49\degr \\
  31  &14 18 37.75 &52 51 28.9 &1.14 &$  0.54 \pm .09 $&		 & \\
  32  &14 18 45.92 &52 51 42.0 &1.05 &$  0.41 \pm .08 $&           & outside IRAC area \\
  33  &14 19 46.17 &52 56 47.2 &1.20 &$  0.83 \pm 0.14$&  0.6 0.0	 & \\
  34  &14 20 50.34 &52 57 46.9 &1.44 &$  1.4  \pm 0.2 $&  0.8 0.0	 & \\
  35  &14 21 01.41 &52 57 55.4 &1.55 &$  0.87 \pm 0.19$&           & outside IRAC area \\
  36  &14 20 33.36 &52 58 00.8 &1.03 &$  0.59 \pm 0.12$&  0.5 0.0	 & \\
  37  &14 20 33.26 &53 00 03.8 &1.18 &$  16.7 \pm 0.16$&           & \\
  38  &14 21 04.93 &53 02 09.4 &1.22 &$  1.64 \pm 0.25$&  1.8 1.1  & \\
  41  &14 20 33.35 &53 08 21.0 &1.05 &$  2.3  \pm 0.14$&  0.8 0.1	 & \\
  44  &14 20 56.85 &53 13 07.5 &1.00 &$  1.0  \pm 0.12$&  1.0 0.4	 & \\
  45  &14 21 54.78 &53 15 00.6 &1.13 &$  4.2  \pm 0.09$&           & \\
  46  &14 21 27.90 &53 15 16.1 &1.18 &$  0.51 \pm 0.1 $&           & \\
  47  &14 21 37.15 &53 20 55.2 &1.01 &$  0.78 \pm 0.17$&  1.5 0.8  &  \\
  48  &14 22 01.41 &53 27 55.3 &1.43 &$  0.80 \pm 0.14$&           & outside IRAC area \\
  49  &14 23 12.71 &53 27 56.7 &1.07 &$  0.91 \pm 0.17$&  1.4 0.0  &  \\
\tableline
\multicolumn{4}{l}{Possible multiple sources}\\
  5   &14 14 53.00 &52 10 25.2 & 1.19&$  1.5  \pm 0.3 $& 3.5 $<$0.6& size uncertain, outside IRAC area\\
  6   &14 14 53.35 &52 10 29.0 & 1.19&$  0.85 \pm 0.18$&           & 5\farcs8 from 5, outside IRAC area\\
  10     &14 16 22.85&52 19 15.9&    &$ 21.7 \pm 0.5  $& 1.8 $<$0.6&classical double morphology, \\
  11     &14 16 22.56&52 19 18.0&    &$ 4.4 \pm 0.3   $& 1.0 $<$1.0&3\farcs1 separation \\
\multicolumn{2}{l}{$10+11$ tot} &                  &1.05&$ 26 \pm 3      $&           &  \\
  17  &14 16 25.55 &52 27 14.4 &     &$  2.5  \pm 0.5 $&  1.2      &3\farcs6 from 18, outside IRAC area \\
  18  &14 16 25.79 &52 27 17.3 &     &$  3.9  \pm 0.5 $&  2.1      & outside IRAC area \\
\multicolumn{2}{l}{$17+18$ tot}&           &1.28 &$  6.5  \pm 1.5 $&           & \\
  26  &14 17 32.84 &52 38 15.0 &1.15 &$  12   \pm 2   $&           & 10\farcs1 from 27 \\
  27  &14 17 32.70 &52 38 25.0 &1.15 &$  5.1  \pm 1.5 $&           & multiple spots \\
  39  &14 21 19.29 &53 03 22.6 &1.13 &$  14.6 \pm 0.15$&           & 12\farcs3 from 40\\
  40  &14 21 18.24 &53 03 30.4 &1.10 &$  16.2 \pm 0.18$&  1.0 0.6  & \\
  42  &14 21 26.99 &53 10 47.0 &1.16 &$  0.64 \pm 0.09$&           & 7\farcs5 from 43 \\
  43  &14 21 26.19 &53 10 49.1 &1.15 &$  2.0  \pm 0.2 $&  1.5 1.1  & \\
  50  &14 22 51.01& 53 36 13.9 &     &$   4.4 \pm 0.3 $&           &classical double morphology, \\
  51  &14 22 51.00& 53 36 19.0 &     &$   6.0 \pm 0.4  $&           &5\farcs5 separation \\
\multicolumn{2}{l}{$50+51$ tot}&           &1.27 &$  12   \pm 2 $&           &  \\

\enddata


\tablecomments{``PB corr'' is the primary beam correction, which has
been applied to the flux densities in column 5.  Sizes of major and
minor axes are given in arcsec after deconvolving the synthesized
beam.  Sources are considered possible multiples if there are two
resolved components within 15$''$.  Combined flux densities are given
only if the two components are potentially blended in the radio
data.  Systematic position uncertainties are 0\farcs1 rms.}

\end{deluxetable}

\begin{deluxetable}{cccc}
\tabletypesize{\footnotesize}
\tablecaption{Comparison with FWKK sources}
\tablewidth{0pt}
\tablehead{
\colhead{FWKK name} &  
\colhead{this paper name} &      
\colhead{$S_5(\rm{FWKK})$} &
\colhead{$S_5(\rm{this~paper})$\tablenotemark{a}} \\
\colhead{15V}&
\colhead{EGS06}&
\colhead{(mJy)}&
\colhead{(mJy)}
}
\startdata
02 & ---   & $0.45 \pm 0.02$ & \0(0.23) \\
03 & ---   & $0.23 \pm 0.03$	& $<$0.2\phm{$<$} \\
10 &  21   & $1.91 \pm 0.07$\tablenotemark{b} & $1.0\pm0.2$ \\
21 & ---   & $0.30 \pm 0.02$	& \0(0.23) \\
34 &  19   & $1.31 \pm 0.03$	& 1.3   \\
50 &  14   & $0.70 \pm 0.01$	& 0.6   \\
70 &  23   & $0.58 \pm 0.02$	& 1.0   \\
78 & ---   & $0.22 \pm 0.04$	& $<$0.3\phm{$<$} \\
\enddata


\tablenotetext{a}{Flux densities in parentheses are below the
  detection threshold of this paper but were measured from the
  mosaics.  Uncertainties in peak flux density are about
  0.06~mJy/beam, and all sources are nearly pointlike at our
  resolution (according to FWKK) {\em except} 15V~10.  Upper limits
  given are 3$\sigma$ assuming point sources.}

\tablenotetext{b}{FWKK give the total flux density of this extended
  source: core and two lobes. (See FWKK Fig.~3a.)  The lobes are
  resolved out in our BnA data, and our flux density includes only
  the core.  Table~4 of FWKK suggests that the core comprises about
  40\% of the total flux density.}

\end{deluxetable}

\begin{deluxetable}{rcccl}
\tabletypesize{\footnotesize}
\tablecaption{IRAC Counterparts to Radio Sources}
\tablehead{
\colhead{EGS06} &  
\colhead{R.A} &      
\colhead{Dec} &
\colhead{sep(\arcsec)} &
\colhead{comments}
}
\startdata
\multicolumn{5}{l}{Isolated radio sources}\\
 1     &14:15:03.89  &+52:01:06.7 & 0.3 &\\
 2     &14:14:33.33  &+52:02:52.9 & 0.3	&interacting component; see Fig.~\ref{f:unid}\\
 3     &14:14:36.76  &+52:05:03.1 & 0.1 &$z=0.655$, old (wk [\ion{O}{2}])\\
 4     &14:15:08.84  &+52:06:42.0 & 0.2	&\\
 7     &14:16:15.07  &+52:11:21.6 & 0.0	&\\
 8     &14:16:11.84  &+52:12:04.0 & 0.8	&\\
13     &14:16:21.86  &+52:25:03.5 & 0.3	&\\
16     &14:16:27.89  &+52:27:07.2 & 0.6	&X-ray\\
14     &14:17:59.30  &+52:25:53.8 & 0.2	&edge of IRAC coverage, poor data\\
19     &14:17:49.21  &+52:28:03.2 & 0.1 &$z=0.996$\tablenotemark{a},
  old (wk [\ion{O}{2}]), X-ray\\
21     &14:17:32.62  &+52:32:03.3 & 0.2 &$z=0.986$, post-SB, X-ray\\
22     &14:17:13.63  &+52:32:13.9 & 0.4 &$z=0.835$, old\\
23     &14:18:09.96  &+52:33:00.4 & 0.2 &\\
24     &14:17:53.40  &+52:35:39.6 & 0.3 &$z=0.679$, old\\
28     &14:18:05.56  &+52:40:32.8 & 0.5 &$z=1.413$, broad-line AGN\\
29     &14:19:45.50  &+52:46:48.0 & 0.1	&\\
30     &14:19:10.43  &+52:48:30.6 & 0.3	&$z=0.346$, old\\
31     &14:18:37.77  &+52:51:28.6 & 0.4	&\\
33     &14:19:46.08  &+52:56:47.1 & 0.8	&\\
34     &14:20:50.37  &+52:57:46.9 & 0.3	&\\
36     &14:20:33.37  &+52:58:00.9 & 0.1	&$z=0.570$, old\\
37     &14:20:33.27  &+53:00:03.7 & 0.1	&\\
41     &14:20:33.34  &+53:08:21.0 & 0.1	&\\
44     &14:20:56.84  &+53:13:07.7 & 0.2	&$z=0.742$, post-SB\\
45     &14:21:54.80  &+53:15:00.6 & 0.2	&\\
46     &14:21:27.87  &+53:15:16.1 & 0.3 &close double in R\\
47     &14:21:37.14  &+53:20:55.1 & 0.1	&\\
49     &14:23:12.67  &+53:27:56.9 & 0.4 &multiple in R\\
\tablebreak
\multicolumn{5}{l}{Possible radio doubles}\\			 
6      &14:14:53.34  &+52:10:28.8 & 0.2 &poor data at 3.6 \& 5.8~\micron\\
$10/11$&14:16:22.76  &+52:19:16.4 &     &$z=1.600$, broad-line AGN, X-ray\\
$26/27$&14:17:32.55  &+52:38:18.1 & 	&4\farcs1 NE of 26\\
$39/40$&14:21:18.42  &+53:03:29.0 &     &2\farcs1 SE of 40,
$z=0.973$, post-SB $+$ narrow-line AGN\\
$42/43$&14:21:26.48  &+53:10:48.7 &     &2\farcs6 W of 43\\
$50/51$&14:22:51.02  &+53:36:16.8 &    	&exposure time 0.1$\times$normal\\
       
\enddata

\tablenotetext{a}{\citet{Hammer1995} found $z=0.838$ and elliptical
  classification for EGS06~19 = 15V~34, but the DEEP2 redshift seems
  secure.  See Appendix~A.}
\tablecomments{Redshifts and spectral classifications are from the
  DEEP2 survey \citep{Davis2003}.  Comment ``X-ray'' indicates
  detection by either Chandra \citep{egs_cxo} or XMM-Newton
  \citep{was03}.}
\end{deluxetable}

\begin{deluxetable}{lrrrrrrrr}
\tablecaption{Photometry\tablenotemark{a}}
\tablehead{
\colhead{EGS06} &  
\colhead{$B$} &  
\colhead{$R$} &      
\colhead{$I$} &
\colhead{3.6~\micron} &
\colhead{4.5~\micron} &
\colhead{5.8~\micron} &
\colhead{8.0~\micron} &
\colhead{24~\micron}
}
\startdata
1       &\no    &\no    &\no    &15.1   &15.9   &12.2   &9.9    &\no   \\
2       &\no    &\no    &\no    &85.8   &68.0   &94.4   &176.2  &{b}   \\
3       &1.15   &9.23   &23.18  &120.4  &79.9   &70.5   &50.4   &54    \\
4       &0.52   &2.33   &4.98   &79.7   &85.1   &99.0   &156.6  &166   \\
7       &0.94   &13.44  &39.45  &165.8  &100.4  &81.7   &45.2   &\no   \\
8       &0.25   &0.31   &0.89   &27.1   &28.3   &19.5   &14.9   &\no   \\
13      &0.28   &0.53   &2.21   &63.8   &64.6   &44.7   &31.9   &\no   \\
16      &1.47   &10.73  &40.59  &132.1  &95.1   &80.8   &80.8   &278    \\
14      &0.57   &1.68   &5.04   &79.4:  &78.3:  &92.5:  &45.4:  &\no   \\
19    &$<$0.3\0 &2.13   &7.88   &111.9  &80.9   &57.3   &40.7   &\no   \\
21      &0.29   &2.67   &8.11   &78.3   &57.5   &41.5   &36.2   &\no   \\
22      &0.51   &7.32   &26.25  &164.9  &106.3  &76.8   &48.8   &\no   \\
23      &\no    &\no    &\no    &17.7   &20.5   &21.5   &28.0   &22      \\
24      &0.70   &10.10  &29.65  &149.4  &88.4   &70.3   &40.1   &\no   \\
28      &1.61   &3.02   &4.51   &42.2   &49.4   &52.8   &65.8   &77      \\
29      &9.55   &21.64  &28.95  &123.4  &148.0  &198.6  &359.1  &598     \\
30      &10.34  &90.12  &171.40 &268.9  &211.1  &130.9  &85.7   &\no   \\
31      &\no    &\no    &\no    &37.7   &42.6   &26.5   &10.2:   &\no   \\
33      &\no    &\no    &\no    &25.9   &30.4   &31.2   &23.8   &29      \\
34      &\no    &\no    &\no    &39.6   &54.1   &70.3   &108.   &139     \\
36      &2.15   &25.89  &59.27  &175.4  &116.9  &88.2   &47.51  &\no   \\
37      &\no    &\no    &\no    &18.8   &23.8   &24.8   &21.3   &225     \\
41      &3.24   &30.39  &102.40 &249.6  &149.7  &126.1  &77.3   &\no   \\
44      &24.39  &48.98  &100.00 &418.0  &252.1  &189.7  &150.2  &27      \\
45      &0.19   &0.35   &2.28   &106.4  &100.6  &71.8   &61.5   &25      \\
46      &0.37   &3.06   &12.85  &134.0  &96.0   &61.8   &43.3   &\no   \\
47      &0.69   &1.70   &3.70   &96.0   &100.1  &71.4   &53.0   &\no   \\
49      &\no    &\no    &\no    &31.1   &36.5   &30.4   &31.2   &57      \\
6       &\no    &46.13  &101.8: &388.0  &293.2  &232.9  &125.1  &\no   \\
10/11   &1.46   &5.73   &14.43  &185.4  &267.9  &365.4  &561.8  &792     \\
26/27   &1.90   &19.12  &53.26  &223.0  &154.7  &138.4  &126.1  &109     \\
39/40   &3.34   &9.38   &29.03  &142.8  &109.2  &86.7   &115.8  &894     \\
42/43   &\no    &\no    &\no    &99.7   &92.9   &61.4   &43.9   &\no   \\
50/51   &7.17   &60.87  &112.20 &177.2  &142.6  &\no    &60.2   &\no   \\
noise\tablenotemark{c}
        &0.08   &0.10   &0.19   &0.2    &0.2    &1.2    &1.2    &18 \\
\enddata

\tablenotetext{a}{Flux densities in $\mu$Jy.  Colon indicates
 photometric  uncertainty $\ga$30\%.}
\tablenotetext{b}{Blended with other sources; see
 Fig.~\ref{f:unid}.}.
\tablenotetext{c}{Typical uncertainty from photon noise alone.  See text for
 discussion of additional measurement uncertainties.}
\end{deluxetable}


\end{document}